\documentclass{article}

\def \be {\begin{equation}}
\def \ee {\end{equation}}
\def \bea {\begin{eqnarray}}
\def \eea {\end{eqnarray}}
\def \nn {\nonumber}

\def \a {\alpha}
\def \b {\beta}

\def \G {\Gamma}
\def \d {\delta}

\def \m {\mu}
\def \n {\nu}
\def \k {\kappa}

\def \s {\sigma}
\def \r {\rho}
\def \o {\omega}

\def \th {\theta}
\def \Th {\Theta}

\def \t {\tau}
\def \dag {\dagger}
\def \p {\partial}

\def\bd{\begin{document}}
\def\ed{\end{document}}
\def\nn{\nonumber}
\def\bea{\begin{eqnarray}}
\def\eea{\end{eqnarray}}
\let\bm=\bibitem
\let\la=\label

\def\N{{\cal N}}
\def\sst{\scriptscriptstyle}
\def\thetabar{\bar\theta}
\def\Tr{{\rm Tr}}
\def\one{\mbox{1 \kern-.59em {\rm l}}}

\def\a{\alpha}      \def\da{{\dot\alpha}}
\def\b{\beta}       \def\db{{\dot\beta}}
\def\c{\gamma}  \def\C{\Gamma}  \def\cdt{\dot\gamma}
\def\d{\delta}  \def\D{\Delta}  \def\ddt{\dot\delta}
\def\e{\epsilon}        \def\vare{\varepsilon}
\def\f{\phi}    \def\F{\Phi}    \def\vvf{\f}
\def\h{\eta}
\def\k{\kappa}
\def\l{\lambda} \def\L{\Lambda}
\def\m{\mu} \def\n{\nu}
\def\o{\omega}
\def\P{\Pi}
\def\r{\rho}
\def\s{\sigma}  \def\S{\Sigma}
\def\t{\tau}
\def\th{\theta} \def\Th{\Theta} \def\vth{\vartheta}
\def\X{\Xeta}
\def\z{\zeta}
\def\w{\wedge}
\def\u{\underline}
\def\hs{\hspace}

\def\cA{{\cal A}} \def\cB{{\cal B}} \def\cC{{\cal C}}
\def\cD{{\cal D}} \def\cE{{\cal E}} \def\cF{{\cal F}}
\def\cG{{\cal G}} \def\cH{{\cal H}} \def\cI{{\cal I}}
\def\cJ{{\cal J}} \def\cK{{\cal K}} \def\cL{{\cal L}}
\def\cM{{\cal M}} \def\cN{{\cal N}} \def\cO{{\cal O}}
\def\cP{{\cal P}} \def\cQ{{\cal Q}} \def\cR{{\cal R}}
\def\cS{{\cal S}} \def\cT{{\cal T}} \def\cU{{\cal U}}
\def\cV{{\cal V}} \def\cW{{\cal W}} \def\cX{{\cal X}}
\def\cY{{\cal Y}} \def\cZ{{\cal Z}}

\def\ua{\underline{\alpha}} \def\ubb{\underline{\beta}}
\def\ug{\underline{\gamma}}
\def\ub{\underline{\phantom{\alpha}}\!\!\!\beta}
\def\uc{\underline{\phantom{\alpha}}\!\!\!\gamma}
\def\um{\underline{\mu}} \def\un{\underline{\nu}}
\def\ud{\underline\delta}
\def\ue{\underline\epsilon}
\def\una{\underline a}\def\unA{\underline A}
\def\unb{\underline b}\def\unB{\underline B}
\def\unc{\underline c}\def\unC{\underline C}
\def\und{\underline d}\def\unD{\underline D}
\def\une{\underline e}\def\unE{\underline E}
\def\unf{\underline{\phantom{e}}\!\!\!\! f}\def\unF{\underline F}
\def\unm{\underline m}\def\unM{\underline M}
\def\unn{\underline n}\def\unN{\underline N}
\def\unp{\underline{\phantom{a}}\!\!\! p}\def\unP{\underline P}
\def\unq{\underline{\phantom{a}}\!\!\! q}
\def\unQ{\underline{\phantom{A}}\!\!\!\! Q}
\def\unH{\underline{H}}
\def\ul{\underline}

\def\As {{A \hspace{-6.4pt} \slash}\;}
\def\bs {{b \hspace{-6.4pt} \slash}\;}
\def\Ds {{D \hspace{-6.4pt} \slash}\;}
\def\ds {{\del \hspace{-6.4pt} \slash}\;}
\def\ss {{\s \hspace{-6.4pt} \slash}\;}
\def\ks {{ k \hspace{-6.4pt} \slash}\;}
\def\ps {{p \hspace{-6.4pt} \slash}\;}
\def\pas {{{p_1} \hspace{-6.4pt} \slash}\;}
\def\pbs {{{p_2} \hspace{-6.4pt} \slash}\;}

\def\Fh{\hat{F}}
\def\Vh{\hat{V}}
\def\Xh{\hat{X}}
\def\ah{\hat{a}}
\def\xh{\hat{x}}
\def\yh{\hat{y}}
\def\ph{\hat{p}}
\def\xih{\hat{\xi}}

\def\psit{\tilde{\psi}}
\def\Psit{\tilde{\Psi}}
\def\tht{\tilde{\th}}

\def\At{\tilde{A}}
\def\Qt{\tilde{Q}}
\def\Rt{\tilde{R}}
\def\Nt{\tilde{N}}

\def\at{\tilde{a}}
\def\st{\tilde{s}}
\def\ft{\tilde{f}}
\def\pt{\tilde{p}}
\def\qt{\tilde{q}}
\def\vt{\tilde{v}}
\def\nt{\tilde{n}}

\def\delb{\bar{\partial}}
\def\bz{\bar{z}}
\def\bD{\bar{D}}
\def\bB{\bar{B}}

\def\bk{{\bf k}}
\def\bl{{\bf l}}
\def\bp{{\bf p}}
\def\bq{{\bf q}}
\def\br{{\bf r}}
\def\bx{{\bf x}}
\def\by{{\bf y}}
\def\bR{{\bf R}}
\def\bV{{\bf V}}

\def\d{\delta}\def\D{\Delta}\def\ddt{\dot\delta}

\def\p{\partial} \def\del{\partial}
\def\xx{\times}
\def\uno{\mbox{1 \kern-.59em {\rm l}}}

\def\trp{^{\top}}
\def\inv{^{-1}}
\def\dag{{^{\dagger}}}

\def\pr{\prime}
\def\rar{\rightarrow}
\def\lar{\leftarrow}
\def\lrar{\leftrightarrow}

\title{\bf Holographical Description of BPS Wilson Loops in Flavored ABJM Theory}
\author{
Bin Chen$^{1, 2, 3}$\footnote{bchen01@pku.edu.cn}, Jun-Bao Wu$^{4}$\footnote{wujb@ihep.ac.cn}~~and Meng-Qi Zhu$^5$\footnote{mzhu@sissa.it}
}
\date{}
\begin{document}
\maketitle
\begin{center}
{\it
$^1$Department of Physics, and State Key Laboratory of Nuclear Physics and Technology, Peking University, No. 5 Yiheyuan Rd, Beijing 100871, P.~R.~China\\
\vspace{2mm}
$^{2}$Collaborative Innovation Center of Quantum Matter, 5 Yiheyuan Rd, \\Beijing 100871, P.~R.~China\\\vspace{2mm}
$^{3}$Center for High Energy Physics, Peking University, 5 Yiheyuan Rd, \\Beijing 100871, P.~R.~China\\
\vspace{2mm}
$^4$Institute of High Energy Physics,
and Theoretical Physics Center for Science Facilities,\\
Chinese Academy of Sciences, 19B Yuquan Road,
Beijing 100049, P.~R.~China\\
\vspace{2mm}
$^5$SISSA, Via Bonomea 265, I 34136, Trieste, Italy\\}


\vspace{10mm}
\end{center}

\date{}
\begin{abstract}
As holographic description of BPS Wilson loops in ${\cal N}=3$ flavored ABJM theory with $N_f=k=1$, BPS M2-branes in $AdS_4\times N(1, 1)$ are studied in details.
Two $1/3$-BPS membrane configurations are found. One of them is dual to the $1/3$-BPS Wilson loop of Gaiotto-Yin type. The regulated membrane action captures precisely the leading exponential behavior of the vacuum expectation values of $1/3$-BPS Wilson loops in the strong coupling limit, which was computed before using supersymmetric localization technique. Moreover, there is no BPS membrane with more supersymmetries in the background, under quite natural assumption on the membrane worldvolume. This suggests that there is no Wilson loop preserving more than 1/3 supersymmetries in such flavored ABJM theory.

 \end{abstract}

\section{Introduction}

Supersymmetric Wilson loop is an important probe  in studying supersymmetric quantum field theory and AdS/CFT correspondence. In the field theory, the computation of the vacuum expectation value(VEV) of such Bogomol'nyi-Prasad-Sommerfield(BPS) Wilson loop often boils down to a matrix model, within recent developments of localization techniques \cite{Pestun:2007rz}. This allows us to obtain exact results to all loops. On the other hand, if the supersymmetric field theory has a holographic dual, the BPS Wilson loop in certain representations could be dual to a fundamental string \cite{Rey, Mal98} or even D-brane \cite{DrukkerFoil, Yamaguchi:2006D5} ending on the contour of the loop \footnote{Recently, an novel and interesting method to obtain certain D-brane solutions dual to Wilson loops appeared in \cite{Fiol:2014vqa}.}. This opens a new window to study the AdS/CFT correspondence and the string interactions.

Among various studies on the BPS Wilson loop, the ones in three-dimensional(3d) supersymmetric Chern-Simons-matter theories are of particular interest. 
The gauge field in  pure Chern-Simons theory is not dynamical, and the study of Wilson loops in this theory leads to important results on Jones polynomials for knots \cite{Witten:1988hf}.
When the Chern-Simons field is coupled to matter, the theory is no longer topological and displays more interesting dynamics.
Moreover the Chern-Simons-matter theories could be the low energy effective action of membranes, and in the large $N$ limit they may dual to M-theory on AdS$_4 \times M_7$ or the reduced type IIA string theory on AdS$_4 \times M_6$.
In \cite{Gaiotto:2007qi}, Gaiotto and Yin constructed the BPS Wilson loops in three-dimensional ${\cal N}=2$ and ${\cal N}=3$ supersymmetric Chern-Simons-matter theories.
The constructions  are similar to the one of half-BPS Wilson loops in four-dimensional ${\cal N}=4$ and ${\cal N}=2$ gauge theories \cite{Rey, Mal98}.
For the $3d$ theories with ${\cal N}=2$ supersymmetries, the Wilson loop is
\be  W=\frac1{\rm dim(R)} {\rm Tr}_R P\exp\left[\int d\tau (iA_{\mu}\dot{x}^\mu+\sigma|\dot{x}|)\right], \label{half}\ee
with $\sigma$ being the auxiliary scalar in the same supermultiplet of the gauge field. It was found to be half-BPS.
For the theories with ${\cal N}=3$ supersymmetries, the Wilson loop is the following $1/3$-BPS one
\be  W=\frac1{\rm dim(R)} {\rm Tr}_R P\exp\left[\int d\tau (iA_{\mu}\dot{x}^\mu+\sum_{a=1}^3\phi_a s^a|\dot{x}|)\right].\ee
Here $s_a$'s are three constants satisfying $\sum (s^a)^2=1$, and $\phi_a$'s are the three auxiliary fields $\Phi_1, \Phi_2, \sigma$ respectively.
Here $\Phi_1, \Phi_2$ come from the (complex) scalar component of an auxiliary chiral superfield in the adjoint representation of the gauge group.
The  BPS Wilson loop of  similar structure in the Aharony-Bergman-Jafferis-Maldacena (ABJM) theory\cite{ABJM}, which is an ${\cal N}=6$ supersymmetric Chern-Simons-matter theory and describes low energy dynamics of the multiple membranes, was constructed in \cite{DPY, CW, RSY}.
In this construction, the auxiliary field $\sigma$ in Eq.~(\ref{half}) was replaced by its on-shell value. After some detailed computations,  this Wilson loop was found to preserve only $1/6$ of the ${\cal N}=6$ supersymmtries in the theory.
The supersymmetry  enhancement  of the action from ${\cal N}=3$ to ${\cal N}=6$  via clever choice of the matter content and the superpotential
does not happen for the Wilson loop of Gaiotto-Yin (GY) type. The ABJM theory has a holographic description in terms of type IIA string theory on $AdS_4\times CP^3$ or M-theory on $AdS_4\times S^7/Z_k$.
In the IIA string description, an F-string with  worldsheet $AdS_2\subset AdS_4$ is believed to be dual to the Wilson loop in the fundamental representation. This  F-string solution was found to preserve half of the supersymmetries of the background \cite{DPY, RSY}. It can also be uplifted to an M2-brane in $AdS_4\times S^7/Z_k$. The worldvolume of this M2-brane is $AdS_2\times S^1$. The $AdS_2$ part is inside $AdS_4$ and $S^1$ is inside $S^7/Z_k$ and along the direction performing the $Z_k$ orbifolding.
The presence of such half-BPS F1 or M2 brane suggests that there should be  half-BPS Wilson loops in the ABJM theory.  Eventually, the sought-after  Wilson loop  was  constructed successfully in \cite{Drukker:2009hy} by including the fermions in bi-fundamental representation into the construction.  To incorporate the fermion, the connection in the GY-type Wilson loop has to be augmented to a superconnection, whose holonomy gives the Wilson loop wanted. By using localization \cite{Kapustin, Drukker:2009hy}, the VEV of both  1/6- and 1/2-BPS Wilson loops can be computed as the correlation functions in the matrix models. It turned out that in the planar and strong coupling limits, the leading exponential behaviors of the VEVs are in perfect agreement with holographic results \cite{Marino:2009jd}, which could be read from the regulated action of the dual macroscopic open string. Such novel construction of BPS Wilson loops was explained via the Brout-Englert-Higgs(BEH) mechanism and was generalized to $2/5$-BPS Wilson loops in ${\cal N}=5$ theories \cite{Hosomichi:2008jb, ABJ} in \cite{LeeLee}.   For other studies on the BPS Wilson loop in the context of the AdS$_4$/CFT$_3$ correspondence, see  \cite{Kim:2012nd}-\cite{Farquet:2014bda}.

The interesting story of the BPS Wilson loops in the ABJM theory leads us to consider whether there are BPS Wilson loops beyond GY-type in less-supersymmetric Chern-Simons-matter theory. In this letter we will focus on the flavored ABJM theory  \cite{Hohenegger:2009as}-\cite{Hikida:2009tp} as a representative of 3d ${\cal N}=3$ theories. In fact,  the ABJM theory could be coupled to $N_f$ flavors in fundamental representations of the gauge groups in a manner preserving ${\cal N}=3$ supersymmetries.  These fundamental flavors are induced by introducing  probe D6-branes in $AdS_4\times CP^3$ appropriately. Taking into account of the backreaction of these D6-branes, the flavored ABJM theory is dual to M-theory  on $AdS_4\times M_7(N_f, N_f, k)$, where the Eschenburg space $M_7(N_f, N_f, k)$ is  a special $3$-Sasakian manifold (only when $N_f$ and $k$ are coprime, the metric on this manifold is smooth)\footnote{The IIA background from the backreaction of these branes was studied in \cite{Conde:2011sw}.}. It is quite difficult to search for half-BPS Wilson loops in the flavored ABJM theory by using the method of \cite{LeeLee}. Inspired by
 the story in the ABJM theory, we first turn to the dual gravitational description. In this case, as the study in the IIA string description is harder than the one in the M-theory setup,
we start with searching for BPS M2-branes in $AdS_4\times M_7(N_f, N_f, k)$. The experience in the ABJM theory leads us to make the ansatz that the embedded M2-brane is of the worldvolume $AdS_2\times S^1$.
The $S^1$ should be along a M-theory circle generated by a supersymmetry-preserving Killing vectors such that the configuration has the chance to preserve the largest amount of supersymmetries.


To find the supersymmetries preserved by the M2-brane,  we first need to find explicitly the Killing spinors on $AdS_4\times M_7$. The metric on $M_7(N_f, N_f, k)$ can be obtained from the supergravity solutions in \cite{GGPT}. However it is very hard to solve the Killing spinor equations in this way. Fortunately, among  $M_7(N_f, N_f, k)$ there is a special one, $M_7(1, 1, 1)\equiv N(1, 1)$, which is just a coset space $SU(3)/U(1)$ \cite{pagepope, Gauntlett:2005jb}. In fact, using this coset description,  the metric of $N(1, 1)$ can be expressed in terms of a coordinate system which is more suitable for solving the Killing spinor equations. After finding the Killing spinors in AdS$_4 \times N(1,1)$, we try to search for the BPS membrane configurations with the ansatz we mentioned above. 
We manage to find two 1/3-BPS membrane configurations. Moreover, we argue that there is no M2-brane preserving more supersymmetries. This indicates that there is no Wilson loop keeping more than 1/3 supersymmetries in the flavored ABJM theory.

M-theory on AdS$_4 \times N(1,1)$ is dual to the ABJM theory with Chern-Simons level $k=1$ coupled to $N_f=1$ flavor\cite{Gaiotto:2009tk, Fujita:2009xz}. The known Wilson loop in this theory is of GY type and keeps one-third supersymmetries. We suggest this 1/3-BPS Wilson loop is dual to one of the membrane configurations we found.  We actually show that the regulated action of the membrane configuration captures precisely the leading exponential behavior of the VEV of the Wilson loop in the strong coupling limit found in \cite{Santamaria:2010dm} via localization techniques.



The remaining parts of the work is organized as follows. We review the field theory results in Sec. 2. And we present the analysis on the Killing spinors in AdS$_4 \times N(1,1)$ in Sec. 3. In Sec. 4, we propose the 1/3-BPS membrane configurations. We end with some discussions in Sec. 5.

\section{Field theory results}

The ABJM theory \cite{ABJM} is a three-dimensional ${\cal N}=6$ Chern-Simons-matter theory. The gauge group of the theory is $U_1(N)\times U_2(N)$
with Chern-Simons levels $(k, -k)$. The matter part includes  ${\cal N}=2$ bifundamental chiral superfields $A_1, A_2$ and anti-bifundamental chiral superfields
$B_1, B_2$. The action of the ABJM theory includes the ${\cal N}=2$  supersymmetric Chern-Simons action,
\be S_{CS}^{{\cal N}=2}=\sum_{i=1}^2  \frac{k_i}{4\pi}\int {\rm Tr}(A_{(i)}\wedge dA_{(i)}+\frac23A_{(i)}^3-\bar{\chi_{(i)}}\chi_{(i)}+2D_{(i)}\sigma_{(i)}),\ee
where $k_1=-k_2=k$ should be an integer. The auxiliary scalars $D_{(i)}$, $\sigma_{(i)}$ and the two-component spinor $\chi_{(i)}$ are ${\cal N}=2$ super-partners of the Chern-Simons gauge fields $A_{(i)}$, so all of them are in the adjoint representation of $U_i(N)$.
The superpotential of the ABJM theory is
\be  {\cal W}_{ABJM}=-\frac{k}{8\pi}{\rm Tr}(\Phi_{(1)}^2-\Phi_{(2)}^2)+\sum_{i=1}^2{\rm Tr}(B_i\Phi_{(1)}A_i)+\sum_{i=1}^2{\rm Tr}(A_i\Phi_{(2)}B_i).\ee
Here $\Phi_{(i)}$ is a ${\cal N}=2$ chiral superfield in the adjoint representation of $U_i(N)$. Notice there are no kinematic terms for $\Phi_{(i)}$.

One interesting generalization of the ABJM theory is to introduce flavors. One may generically introduce $n_1$ fundamentals of $U_1(N)$, $Q_1^t, \tilde{Q}_{1t}, t=1, \cdots, n_1$ and $n_2$ fundamentals of $U_2(N)$, $Q_2^s, \tilde{Q}_{2s}, s=1, \cdots, n_2$ into the theory but still keep ${\cal N}=3$ supersymmetries \cite{Hohenegger:2009as}-\cite{Hikida:2009tp}. This requires  the contribution of these matters to the superpotential to be
\be  \Delta {\cal W}_{flavors}=-\frac{k}{8\pi}(\sum_{t=1}^{n_1}{\rm Tr}(\tilde{Q}_{1t}\Phi_{(1)}Q_1^t)+\sum_{s=1}^{n_2}{\rm Tr}(\tilde{Q}_{2s}\Phi_{(1)}Q_2^s)).\ee
The superpotential of the full theory is
\be {\cal W}={\cal W}_{ABJM}+\Delta {\cal W}_{flavors}.\ee
  The total numbers of flavors is $N_f\equiv n_1+n_2$. The above construction is motivated by introducing flavor D-branes. In this case, the  ${\cal N}=3$ flavored ABJM theory in the large $N$ limit could be dual to IIA theory with $N_f$ D6-branes wrapping $RP^3$ in $CP^3$, or dual to M-theory on AdS$_4\times M_7(N_f, N_f, k)$.
 For the case with $N_f=k$, $M_7(k, k, k)$  becomes a certain orbifolds $N(1, 1)/Z_k$ for $k>1$. For $N_f=k=1$, this $M_7$ is just $N(1, 1)$.

Let us consider the following Wilson loops \cite{Gaiotto:2007qi} in this theory
\be  W_i=\frac1{\rm dim(R)} {\rm Tr}_R P\exp\left[\int d\tau (iA_{(i)\mu}\dot{x}^\mu+\sum_{a=1}^3\phi_{(i)a} s^a|\dot{x}|)\right],\ee
where 
$s_a$ are three constants satisfying $\sum (s^a)^2=1$, and for each $i=1, 2$, $\phi_{(i)a}$
are the three auxiliary fields as introduced in the introduction.
When the Wilson loop is along a straight line or a circular loop, one-third of the supersymmetries is preserved.
By using the $SO(3)$ R-symmetry of this ${\cal N}=3$ Chern-Simons-matter theory, we will focus on the case with $s^1=s^2=0, s^3=1$, without loss of generality
\footnote{Similar idea was used for the construction of $1/6$-BPS Wilson loops in ABJM theory \cite{DPY, CW, RSY}.}. And now the Wilson loop in the fundamental representation becomes
\be \label{wl}  W_i=\frac1{\rm dim(R)} {\rm Tr}_R P\exp\left[\int d\tau (iA_{(i)\mu}\dot{x}^\mu+\sigma |\dot{x}|)\right].\ee

In the flavored ABJM theory, the strong coupling limit of the VEV of this $1/3$-BPS Wilson loop in the fundamental representation
was computed in \cite{Santamaria:2010dm} based on the supersymmetric localization \cite{Kapustin}. When $n_1=N_f, n_2=0$, the leading exponential behavior of the VEV is
\be <W_i>\sim \exp\left[2\pi \sqrt{\frac{N}{2k+N_f}}\right].\label{field1}\ee
For the special case with $n_1=N_f=k, n_2=0$, we get
\be <W_i>\sim \exp\left[2\pi \sqrt{\frac{N}{3k}}\right].\label{field2}\ee

\section{Background and Killing spinors}
The  metric of the background $AdS_4\times N(1, 1)$ is
\be  ds^2=R^2(\frac14ds_{AdS_4}^2+ds^2_{N(1, 1)}), \label{metric}\ee
with
\be ds_{AdS_4}^2=\cosh^2u(-\cosh^2\rho dt^2+d\rho^2)+du^2+\sinh^2u d\phi^2,\ee
and
\bea  ds^2_{N(1, 1)}&=&\frac12(d\alpha^2+\frac14\sin^2\alpha(\sigma_1^2+\sigma_2^2)
+\frac14\sin^2\alpha\cos^2\alpha\sigma_3^2+\frac12(\Sigma_1-\cos\alpha\sigma_1)^2\nn\\
&+&\frac12(\Sigma_2-\cos\alpha\sigma_2)^2+\frac{1}{2}(\Sigma_3-\frac12(1+\cos^2\alpha)
\sigma_3)^2), \label{n11} \eea
where   $\sigma_i$ and $\Sigma_i$ are right invariant one-forms on $SO(3)$ and $SU(2)$ respectively
\bea \sigma_1&=&\sin\phi_1d\theta_1-\cos\phi_1\sin\theta_1d\psi_1,\\
           \sigma_2&=&\cos\phi_1d\theta_1+\sin\phi_1\sin\theta_1d\psi_1,\\
           \sigma_3&=&d\phi_1+\cos\theta_1d\psi_1,\\
           \Sigma_1&=&\sin\phi_2d\theta_2-\cos\phi_2\sin\theta_2d\psi_2,\\
           \Sigma_2&=&\cos\phi_2d\theta_2+\sin\phi_2\sin\theta_2d\psi_2,\\
           \Sigma_3&=&d\phi_2+\cos\theta_2d\psi_2.
\eea
The ranges of the coordinates are respectively
\bea && \alpha\in[0, \pi/2],\hspace{3ex} \theta_1\in[0, \pi], \hspace{3ex} \phi_1\in[0, 4\pi],\hspace{3ex} \psi_1\in[0, 2\pi], \nonumber\\
&& \theta_2\in [0, \pi],\hspace{3ex}
\phi_2\in[0, 2\pi],\hspace{3ex} \psi_2\in[0, 2\pi].  \eea
The volume of $N(1, 1)$ of unit radius with the metric Eq.~(\ref{n11}) is
\be \mbox{vol}(N(1, 1))=\frac{\pi^4}{8}. \ee
Now the flux quantization gives
\bea R=2\pi l_p\left(\frac{N}{6\cdot\mbox{vol}(N(1, 1))}\right)^{1/6}=l_p\left(\frac{2^8\pi^2N}{3}\right)^{1/6}.\label{rlp}\eea

The background four-form field strength is\footnote{Only for this choice of sign of the field, there are nonzero solutions to the Killing spinor equations \cite{pagepope}.}
\be H=-\frac38 R^3\cosh^2u\cosh\rho \sinh udt\wedge d\rho \wedge du \wedge d\phi.  \ee
Corresponding to the metric (\ref{metric}), the vielbeins could be chosen to be
\bea e^{\ul{0}}&=&\frac{R}{2}\cosh u\cosh\rho dt,\hspace{3ex} e^{\ul{1}}=\frac{R}{2}\cosh ud\rho,\label{frame1}\\
     e^{\ul{2}}&=&\frac{R}{2}du,\hspace{3ex}
     e^{\ul{3}}=\frac{R}2\sinh u d\phi,\\
     e^{\ul{4}}&=&\frac{R}{\sqrt{2}}d\alpha,\hspace{3ex}
     e^{\ul{5}}=\frac{R}{2\sqrt{2}}\sin\alpha\sigma_1,\\
     e^{\ul{6}}&=&\frac{R}{2\sqrt{2}}\sin\alpha\sigma_2,\hspace{3ex}
     e^{\ul{7}}=\frac{R}{2\sqrt{2}}\sin\alpha\cos\alpha\sigma_3,\\
     e^{\ul{8}}&=&\frac{R}{2}(\Sigma_1-\cos\alpha\sigma_1),\hspace{3ex}
     e^{\ul{9}}=\frac{R}{2}(\Sigma_2-\cos\alpha\sigma_2),\\
     e^{\ul{\sharp}}&=&\frac{R}{2}(\Sigma_3-\frac12(1+\cos^2\alpha)\sigma_3).\label{frame11}
      \eea
The  spin connections with respect to  these vielbeins are listed in the Appendix.
In terms of the vielbeins,  the four-form field strength can be written as
\be H=-\frac{6}{R}e^{\ul{0}}\wedge e^{\ul{1}}\wedge e^{\ul{2}}\wedge e^{\ul{3}}.\label{H4}\ee

The Killing spinor equation in  $AdS_4\times N(1, 1)$ is \be \nabla_{\ul{M}}\eta+\frac1{576}(3\Gamma_{{\ul N}{\ul P}{\ul Q}{\ul R}}\Gamma_{\ul M}-\Gamma_{\ul M}\Gamma_{{\ul N}{\ul P}{\ul Q}{\ul R}})H^{{\ul N}{\ul P}{\ul Q}{\ul R}}\eta=0. \ee
It could be cast into the form
\be\nabla_{\ul{M}}\eta-\frac{1}{4R}(3\hat{\Gamma}\Gamma_{\ul{M}}-\Gamma_{\ul{M}}\hat{\Gamma})\eta=0, \ee
with
\be \hat{\Gamma}\equiv\Gamma_{\ul{0123}}.\ee
In this work, our convention about the product of eleven $\Gamma$ matrices is \be\Gamma_{\ul{0}\ul{1}\ul{2}\ul{3}\ul{4}\ul{5}\ul{6}\ul{7}\ul{8}\ul{9}\ul{\sharp}}=1.\ee

For the components in $AdS_4$, i.e., $M=\mu=0, 1, 2 ,3$, we get
\be \nabla_{\ul{\mu}}\eta+\frac{1}{R}\Gamma_{\ul{\mu}}\hat{\Gamma}\eta=0, \label{kseads}\ee
while for the components in $N(1, 1)$, i.e., $M=m=4, \cdots, \sharp$, we have
\be \nabla_{\ul{m}}\eta-\frac{1}{2R}\Gamma_{\ul{m}}\hat{\Gamma}\eta=0.\label{ksen11}\ee
The integrability condition of Eq.~(\ref{ksen11}) gives
\be C^{\ul{abcd}}\Gamma_{\ul{ab}}\eta=0, \ee
where $C^{\ul{abcd}}$ is the Weyl tensor of $N(1, 1)$. After some computations, we find that for the coordinate system and the vielbeins we used,
the above conditions just gives the projection condition \be\Gamma_{\ul{4567}}\eta=-\eta. \label{projection1}\ee

The solutions of the above Killing spinor equations (\ref{kseads}-\ref{ksen11}) are
\bea \eta=e^{\frac{\phi_2}{2}(\Gamma_{\ul{47}}+\Gamma_{\ul{89}})} e^{\frac{\theta_2}{2}(\Gamma_{\ul{46}}-\Gamma_{\ul{8\sharp}})} e^{\frac{\psi_2}{2}(\Gamma_{\ul{47}}+\Gamma_{\ul{89}})} e^{-\frac{u}2\Gamma_{\ul{2}}\hat{\Gamma}}
e^{-\frac{\rho}2\Gamma_{\ul{1}}\hat{\Gamma}}e^{-\frac{t}2\Gamma_{\ul{0}}\hat{\Gamma}}e^{\frac{\phi}{2}\Gamma_{\ul{23}}}\eta_0,\eea
with $\eta_0$ satisfying the following projection conditions
\be \Gamma_{\ul{4567}}\eta_0=-\eta_0, \hs{3ex} (\Gamma_{\ul{58}}+\Gamma_{\ul{69}}+\Gamma_{\ul{7\sharp}}-\Gamma_{\ul{4}}\hat{\Gamma})\eta_0=0. \label{proj2} \ee
The first condition comes from Eq.~(\ref{projection1}), while the second is the additional one appeared in solving the Killing spinor equations.
From these projection conditions, we can easily find that the dimension of the solution space of the Killing spinor equations is\footnote{The number of Killing spinors was obtained in \cite{pagepope}. They did not give the explicit expressions of the Killing spinors which are essential for our study.} $12$. These 12 Killing spinors are dual to  $6$ super-Poincare charges and $6$ superconformal charges in the corresponding three-dimensional ${\cal N}=3$ superconformal field theory.


\section{BPS M2-branes}

As we stressed in the Introduction, the Wilson loops in the fundamental representation should dual to M2-brane with topology $AdS_2\times S^1$ where $AdS_2$ is embedded into $AdS_4$ and $S^1$ is  an M-theory circle in $N(1, 1)$ \cite{Sparks}. The tangent vector of  the M-theory circle should be a supersymmetry-preserving Killing vector $\hat K$.
This means that $\hat K$ should satisfy the following condition
\be {\cal L}_{\hat K}\eta\equiv {\hat K}^{\ul{M}}\nabla_{\ul{M}}\eta+\frac14(\nabla_{\ul{M}}{\hat K}_{\ul{N}})\Gamma^{\ul{M}\ul{N}}\eta=0,\ee for any Killing spinor $\eta$
given in the previous section.
After some computations, we find that the following Killing vectors
\bea  
{\hat K}_1&=&\partial_{\psi_1},\\
{\hat K}_2&=&\partial_{\phi_1}+\partial_{\phi_2}, \eea
which satisfy the above conditions\footnote{Actually there are other supersymmetry-preserving Killing vectors, however they cannot generate a circle.}.

Now we start our search for BPS $M2$-branes in $AdS_4\times N(1, 1)$.
The bosonic part of the $M2$-brane action is:
\be
S_{M2}=T_2\left(\int d^3\xi\sqrt{-\mbox{det}g_{mn}}-\int P[C_3]\right), \ee
where $g_{mn}$ is the induced metric on the membrane, $T_2$ is
the tension of the $M2$-brane: \be T_2={1\over (2\pi)^2l_p^3},\ee and
$P[C_3]$ is the pullback of the bulk 3-form gauge
potential to the worldvolume of the membrane.
The gauge choice for the background 3-form gauge potential  $C_3$ in the case at hand is
\be C_3=\frac{R^3}8(\cosh^3u-1)\cosh\rho dt\w d\rho\w d\phi.
\ee
From the action, the
membrane equation of motion is\footnote{We always use the indices from the beginning (middle) of the alphabet to refer
to the frame (coordinate) indices, and the underlined indices to
refer to the target space ones.} \bea
&&\frac{1}{\sqrt{-g}}\p_m\left(\sqrt{-g}g^{mn}\p_nX^{\underline{N}}\right)G_{\underline{MN}}
+g^{mn}\p_{m}X^{\underline{N}}\p_{n}X^{\underline{P}}\G^{\underline{Q}}_{\underline{NP}}G_{\underline{QM}}\nn\\
&=&\frac{1}{3!\sqrt{-g}}\epsilon^{mnp}H_{\ul{M}\ul{N}\ul{P}\ul{Q}}\p_mX^{\ul{N}}\p_nX^{\ul{P}}\p_pX^{\ul{Q}}.\label{eom}\eea
Notice that $\epsilon^{mnp}$ is a tensor density on the world-volume of the membrane.

We are mainly interested in the BPS $M2$-branes. The supercharges preserved by the $M2$-brane are determined by the following equation
\be \Gamma_{M2}\eta=\eta,\label{m2}\ee
with \be\Gamma_{M2}= \frac{1}{\sqrt{-g}}\partial_\tau X^{\ul{M}}\partial_\xi X^{\ul{N}}\partial_\sigma X^{\ul{P}}e^{\ul{A}}_{\ul{M}}e^{\ul{B}}_{\ul{N}}e^{\ul{C}}_{\ul{P}}\Gamma_{\ul{A}\ul{B}\ul{C}}, \ee
where $\tau, \xi, \sigma$ are coordinates on the worldvolume of the $M2$-brane.

The first ansatz  for M2-brane is
\be t=\tau,\hs{2ex} \rho=\xi,\hs{2ex} \psi_1=\sigma,\hs{2ex} \sigma\in [0, 2\pi].\ee
In other words, the $S^1$ is generated by the Killing vector ${\hat K}_1$.
The induced metric is
\bea d{\tilde s}^2&=&R^2(\frac14\cosh^2u(-\cosh^2\rho d\tau^2+d\xi^2)\nonumber\\
&+&\frac1{256}(45+20\cos2\alpha-\cos4\alpha-8 \cos2\theta_1\sin^4\alpha)d\sigma^2).\eea
And now the $M2$-brane action is
\bea S_{M2}&=&\frac{T_{M2}R^3}{4}\int d^3\sigma \cosh^2u\cosh\rho \nn\\
&\times&\left[ \frac1{256}(45+20\cos2\alpha-\cos4\alpha-8 \cos2\theta_1\sin^4\alpha)\right]^{1/2},\label{action1}\eea
The equations of motion give the constraints that
\bea (u, \alpha)=(0, 0), \eea
or \bea  (u, \alpha, \theta_1)=(0, \frac{\pi}2, 0),\eea
or \bea (u, \alpha, \theta_1)=(0, \frac{\pi}2, \frac{\pi}{2}).\eea

To compute the on-shell action of the $M2$-brane whose boundary at infinity is an $S^1$, we switch to the Euclidean $AdS_4$ with the following metric suitable for circular Wilson loop:
\be ds^2_4=\frac14(\cosh^2u(d\rho^2+\sinh^2\rho d\psi^2)+du^2+\sinh^2ud\phi^2).\label{eads}\ee
The on-shell action of the $M2$-brane, Eq.~(\ref{action1}), becomes
\bea S_{M2}&=&\frac{T_{M2}R^3}{4}\int d\Omega_{EAdS_2} d\sigma \nn\\
& &\left[\frac1{128}(45+20\cos2\alpha-\cos4\alpha-8 \cos2\theta_1\sin^4\alpha)\right]^{1/2},\eea
with \be \int d\Omega_{EAdS_2}=\int  d\rho d\psi\sinh \rho.\ee
Using the fact that $\sigma\in[0, 2\pi]$, $T_{M2}=1/(4\pi^2l_p^3)$ and Eq.~(\ref{rlp}), we find
\bea S_{M2}&=&2\sqrt{\frac{N}{3}\left(\frac1{256}(45+20\cos2\alpha-\cos4\alpha-8 \cos2\theta_1\sin^4\alpha)\right)}\nn\\
&\times&\int d\Omega_{EAdS_2}.\eea
After adding boundary terms to regulate the action as in \cite{DrukkerGrossOoguri}, we get
\bea S_{M2}=-4\pi\sqrt{\frac{N}{3}\left(\frac1{256}(45+20\cos2\alpha-\cos4\alpha-8 \cos2\theta_1\sin^4\alpha)\right)}.\eea
For the M2-brane put at $\alpha=0$, $(\alpha, \theta_1)=(\pi/2, 0)$, or $(\alpha, \theta_1)=(\pi/2, \pi/2)$, the on-shell action is repsectively
\be -2\pi\sqrt{\frac{N}{3}},\hs{2ex} -{\pi}\sqrt{\frac{N}{3}}, \hs{2ex} -{\pi}\sqrt{\frac{2N}{3}}.\label{onshellaction} \ee

We can determine how many supersymmetries are preserved by the configurations. Now we have
\bea\Gamma_{M2}=\frac1{\sqrt{-g}}\frac{R^2}{4}\cosh^2u\cosh\rho \Gamma_{\ul{01}}\tilde\Gamma,\eea
with \bea\tilde\Gamma&=&-\frac{R}{2\sqrt{2}}\sin\alpha\cos\phi_1\sin\theta_1\Gamma_{\ul{5}}+
\frac{R}{2\sqrt{2}}\sin\alpha\sin\phi_1\sin\theta_1\Gamma_{\ul{6}}\nonumber\\
&+&\frac{R}{2\sqrt{2}}\sin\alpha\cos\alpha\cos\theta_1\Gamma_{\ul{7}}+\frac{R}2\cos\alpha\cos\phi_1\sin\theta_1\Gamma_{\ul{8}}\nonumber
\\&-&\frac{R}{2}\cos\alpha\sin\phi_1\sin\theta_1\Gamma_{\ul{9}}-\frac{R}{4}(1+\cos^2\alpha)
\cos\theta_1\Gamma_{\ul{\sharp}}.\eea
Then we find that only the M2 branes put at
\be (u, \alpha, \theta_1, \theta_2)=(0, 0, 0, 0),\ee  or
  \be  (u, \alpha, \theta_1, \theta_2)=(0, \pi/2, 0, 0)\ee
are BPS. Actually they are all $1/3$-BPS. Among them,
 the 1/3-BPS M2-brane put at
\be (u, \alpha, \theta_1, \theta_2)=(0, 0, 0, 0),\ee  gives the dominant contributions to the VEV of the $1/3$-BPS Wilson loops. From Eq.~(\ref{onshellaction}), the holographic prediction for the VEV of this loop
is \be<W>\sim \exp(2\pi\sqrt{\frac{N}{3}}) \ee in the large $N$ limit.  
The other $1/3$-BPS M2-brane solution put at
\be  (u, \alpha, \theta_1, \theta_2)=(0, \pi/2, 0, 0)\ee
may corresponding to sub-leading saddle point in the matrix integral obtained via supersymmetric localization.
Similar phenomenon was found for the F-string in $AdS_5\times S^5$ dual to  $1/4$-BPS Wilson loops in the fundamental (symmetric) representation in ${\cal N}=4$ super Yang-Mills \cite{Drukker:2006ga, Drukker:2006zk}.

The second ansatz for M2-brane is
\be  t=\tau,\hs{2ex} \rho=\xi, \hs{2ex}\phi_1=2\sigma,\hs{2ex} \phi_2=\phi_0+2\sigma,\ee
with $\sigma\in [0, 2\pi]$ and $\phi_0$ a constant. This corresponds to the case that $S^1$ is generated by ${\hat K}_2$. The induced metric is
\be d\tilde{s}^2=R^2(\frac14 \cosh^2u(-\cosh^2\rho d\tau^2+d\xi^2)+\frac1{8}\sin^2\alpha(3+\cos(2\alpha))d\sigma^2). \ee
The equations of motion require\footnote{Notice if we set $\alpha=0$, the $\sigma$ direction will shrink.}
\be u=0, \hs{3ex}\alpha=\pi/2. \ee
 Similar to the above discussion, we find that  the regulated on-shell action to be \be\ S=-2\pi\sqrt{\frac{N}{3}}.\ee
In this case we have
\bea\Gamma_{M2}=\frac1{\sqrt{-g}}\frac{R^2}{4}\cosh^2u\cosh\rho \Gamma_{\ul{01}}\tilde\Gamma,\eea
with \bea\tilde\Gamma&=&\frac{R}{\sqrt{2}}\sin\alpha\cos\alpha\Gamma_{\ul{7}}+\frac{R}{2}\sin^2\alpha\Gamma_{\ul{\sharp}}.\eea
From this we can find that only the M2-brane put at $(u, \alpha, \theta_2)=(0, \pi/2, 0)$ is  BPS, and it is 1/3-BPS.
 The prediction for the VEV of the dual $1/3$-BPS Wilson loop is \be <W>\sim \exp(2\pi\sqrt{\frac{N}{3}}) \ee in the large $N$ limit.

 If we consider M-theory on $AdS_4\times N(1, 1)/Z_k$ with $Z_k$ along ${\hat K}_1$ or ${\hat K}_2$ direction, the flux quantization now gives
  \bea R=2\pi l_p\left(\frac{N}{6\mbox{vol}(N(1, 1)/Z_k)}\right)^{1/6}=l_p\left(\frac{2^8\pi^2Nk}{3}\right)^{1/6},\eea
 and the length of the $\sigma$ direction of the M2-brane worldvolume is reduced by a factor $1/k$. Taking these two effects into account,
 the holographic prediction for the leading exponential behavior of  the VEV of the dual  Wilson loop becomes \be <W>\sim \exp(2\pi\sqrt{\frac{N}{3k}}), \ee
 in either case. This prediction matches exactly with the result derived from the supersymmetric localization method in Eq.~(\ref{field2}).

Finally, we would like to provide another argument to support  our speculation that there is no M2-brane with more supersymmetries than 1/3-BPS. Let us
stress again that the M2-brane has worldvolume $AdS_2\times S^1$ with $AdS_2\subset AdS_4, S^1\subset N(1, 1)$. We can decompose the Killing spinors in $AdS_4\times N(1, 1)$ as
\be  \eta=\epsilon\otimes \alpha,\ee
with $\epsilon$ being a 4-component Killing spinor in $AdS_4$ and $\alpha$ being an 8-component Killing spinor in $N(1, 1)$.  There are projection conditions on $\alpha$ similar to the ones in Eq.~(\ref{proj2}). These projection conditions leave only 3 linear independent components of $\alpha$. Now we turn to the projection condition for BPS M2-brane, Eq.~(\ref{m2}). This equation can be decomposed as
\be \Gamma^{AdS_4}_{M2}\epsilon_{\pm}=\pm\epsilon_{\pm}, \hs{2ex} \Gamma^{N(1,1)}_{M2}\alpha_\pm=\pm\alpha_\pm. \ee
The solution of Eq.~(\ref{m2}) can be either $\epsilon_+\otimes \alpha_+$ or $\epsilon_-\otimes \alpha_-$.
For either choice of the sign, $\Gamma^{AdS_4}_{M2}$ kills half components of $\epsilon$. And the dimensions of the solution space of $\alpha_+$ and $\alpha_-$
should be the same. These two facts lead to the result that the dimension of the solution space of eq.~(\ref{m2}) should be $4n$ with $n$ an integer.  This has ruled out  the existence the membranes preserving $6$ supercharges, i. e., being half-BPS. The probe membrane is not believed to be able to preserve more than half of the supersymmetries of the background. It is reasonable to expect that this argument is also valid for general 3-Sasakian manifolds.
For M-theory on $AdS_4\times S^7/Z_k$, half-BPS means $12$ supercharges. So such membrane is permissible, consistent with the results in \cite{DPY, RSY}.

\section{Conclusion and discussions}

In this paper, we discussed the holographic dual of BPS Wilson loop operators in an ${\cal N}=3$ Chern-Simons-matter field theory. The field theory in our study is dual to M-theory on $AdS_4\times N(1, 1)$ so that the  object dual to the Wilson loop should be a membrane keeping the same amount of supersymmetries. We found the 1/3-BPS membrane configurations, after careful analysis of the Killing spinors in $AdS_4\times N(1, 1)$. We suggested that one of the membrane configurations should be dual to the Wilson loop, and supported the picture by showing the regulated action of the membrane is exactly consistent with the strong coupling behavior of the VEV of the Wilson loop. More importantly, under quite natural ansatz, we found no membrane configuration keeping more supersymmetries in $AdS_4\times N(1, 1)$. This suggest that there could be no Wilson loop  with more supersymmmetries in the flavored ABJM theory.

We would caution the reader that our analysis is on the membrane configurations in  $AdS_4\times N(1, 1)$, which dual to the flavored ABJM theory with only one flavor. We worked under the ansatz that the worldvolume of the membrane is AdS$_2 \times S^1$, with AdS$_2$ being in AdS$_4$ and $S^1$ being along a circle generated by the SUSY-preserving Killing vector. The ansatz is reasonable in the sense that in IIA picture, the worldsheet of F-string is AdS$_2$ being in AdS$_4$, and the uplift to M-theory should keep maximally possible supersymmetries. However, for the flavored ABJM theory with more flavors, the dual geometry includes a more complicated Eschenburg space $M_7(N_f,N_f,k)$, on which the Killing spinor analysis is much more difficult. We are short of direct analysis of BPS membrane configuration in this background, even though we suspect that there would be no membrane with more than 1/3 supersymmetries.

The VEV of the BPS Wilson loop in the flavored ABJM theory has been obtained holographically in \cite{Santamaria:2010dm} using the estimate of AdS radius in type IIA string theory \cite{Gaiotto:2009tk}. The prediction is \be \log(<W>)\sim c\sqrt{\frac{N}{3k}}.\ee
  Here $c$ being an ${\cal O}(1)$ factor  has not been determined rigorously in IIA F-string description. The result was obtained under the assumption that the contributions from the strings ending on D6-branes in the IIA  background  are vanishing or subleading. In this work, we determined that the constant $c=2\pi$ in the membrane description. Moreover, the supersymmetries preserved by the F-string configuration has not been studied in type IIA string theory description, either. Naively one might expect that the F-string can keep half of the supersymmetries. However, the presence of D6-brane make things subtle. It would be interesting to do such analysis directly in IIA string description. We expect that it would give consistent picture with our result using membrane.

  One unsolved problem in our study is to determine which 1/3-BPS membrane corresponds to the 1/3-BPS Wilson loop of GY-type in the field theory. Another related question is that what kind of loop operator corresponds to the other membrane. It could be another  1/3-BPS Wilson loop beyond GY-type, or other type of loop defect. It could also possible to dual to Wilson loops in another field theory dual of M-theory on $AdS_4\times N(1, 1)$.\footnote{We do not expect it is dual to vortex loop in field theory, since the $S^1$ part is completely inside $N(1, 1)$.}  It would be nice to understand these issues better.


Our studies on the M-theory side lead us to conjecture that there are no BPS Wilson loops preserving more than 1/3 supersymmetries  in ${\cal N}=3$ Chern-Simons-matter theories.
Our preliminary studies in ${\cal N}=4$ theories give strong hints that there are half-BPS Wilson loops in such theories \cite{Hosomichi:2008jd, Benna:2008zy, Imamura:2008nn}.
These loops are certainly beyond GY type. We leave the construction and studies of such loop operators for further works.

\section*{Acknowledgments}
We would like to thank  Mitsutoshi Fujita,  Daniel Louis Jafferis and Sangmin Lee for very helpful discussions. This work was in part supported by NSFC Grant No.~11275010(B.~C.), No.~11335012(B.~C.), No.~11325522(B.~C.), No.~11105154(J.~W.), and No.~11222549(J. W.). J.~W. also gratefully acknowledges the support of K.~C.~Wong Education Foundation and Youth Innovation Promotion Association of CAS. M.~Z. would like to thank IHEP, CAS for hospitality.

\section*{Appendix: Spin connection of $AdS_4\times N(1, 1)$}
The  spin connections with respect to vielbeins of $AdS_4\times N(1, 1)$ given in eqs.~(\ref{frame1}-\ref{frame11}) are
 \bea
 \omega_{\ul{0}}^{\ul{0}\ul{1}}&=&\frac{2}{R}\frac{\tanh\rho}{\cosh u},\,
  \omega_{\ul{0}}^{\ul{0}\ul{2}}=\frac{2}{R}\tanh u,\\
  \omega_{\ul{1}}^{\ul{1}\ul{2}}&=&\frac{2}{R}\tanh u,\,
  \omega_{\ul{3}}^{\ul{2}\ul{3}}=-\frac{2}{R} \coth u,\\
  \omega_{\ul{4}}^{\ul{5}\ul{8}}&=&\frac1R,\,\,\,
  \omega_{\ul{4}}^{\ul{6}\ul{9}}=\frac1R,\\
 \omega_{\ul{4}}^{\ul{7}\ul{\sharp}}&=&\frac1R,\,\,\,
 \omega_{\ul{5}}^{\ul{4}\ul{5}}=-\frac{\sqrt{2}}R\cot\alpha,\\
 \omega_{\ul{5}}^{\ul{4}\ul{8}}&=&-\frac1R,\,\,\,
 \omega_{\ul{5}}^{\ul{6}\ul{7}}=-\frac{\sqrt{2}}R\cot\alpha,\\
 \omega_{\ul{5}}^{\ul{6}\ul{\sharp}}&=&\frac1R,\,\,\,
 \omega_{\ul{5}}^{\ul{7}\ul{9}}=-\frac1R,\\
 \omega_{\ul{5}}^{\ul{9}\ul{\sharp}}&=&-\frac{2\sqrt{2}}R\cot\alpha,\,\,\,
\omega_{\ul{6}}^{\ul{4}\ul{6}}=-\frac{\sqrt{2}}R\cot\alpha,\\
 \omega_{\ul{6}}^{\ul{4}\ul{9}}&=&-\frac1R,\,\,\,
 \omega_{\ul{6}}^{\ul{5}\ul{7}}=\frac{\sqrt{2}}{R}\cot\alpha,\\
 \omega_{\ul{6}}^{\ul{5}\ul{\sharp}}&=&-\frac1R,\,\,\,
 \omega_{\ul{6}}^{\ul{7}\ul{8}}=\frac{1}{R},\\
 \omega_{\ul{6}}^{\ul{8}\ul{\sharp}}&=&\frac{2\sqrt{2}}{R}\cot\alpha,\,\,\,
 \omega_{\ul{7}}^{\ul{4}\ul{7}}=-\frac{\sqrt{2}}{R}\cos2\alpha\csc\alpha\sec\alpha,\\
 \omega_{\ul{7}}^{\ul{4}\ul{\sharp}}&=&-\frac1R,\,\,\,
 \omega_{\ul{7}}^{\ul{5}\ul{6}}=\frac{1}{\sqrt{2}R}(-3+\cos2\alpha)\csc\alpha\sec\alpha,\\
 \omega_{\ul{7}}^{\ul{5}\ul{9}}&=&\frac{1}{R},\,\,\,
 \omega_{\ul{7}}^{\ul{6}\ul{8}}=-\frac{1}{R},\\
\omega_{\ul{7}}^{\ul{8}\ul{9}}&=&-\frac{1}{\sqrt{2}R}(3+\cos2\alpha)\csc\alpha\sec\alpha,\,\,\,
\omega_{\ul{8}}^{\ul{4}\ul{5}}=-\frac1R,\\
\omega_{\ul{8}}^{\ul{6}\ul{7}}&=&-\frac1R,\,\,\,
\omega_{\ul{8}}^{\ul{9}\ul{\sharp}}=-\frac1R,\\
\omega_{\ul{9}}^{\ul{4}\ul{6}}&=&-\frac1R,\,\,\,
\omega_{\ul{9}}^{\ul{5}\ul{7}}=\frac1R,\\
\omega_{\ul{9}}^{\ul{8}\ul{\sharp}}&=&\frac1R,\,\,\,
\omega_{\ul{\sharp}}^{\ul{4}\ul{7}}=-\frac1R,\\
\omega_{\ul{\sharp}}^{\ul{5}\ul{6}}&=&-\frac1R,\,\,\,
\omega_{\ul{\sharp}}^{\ul{8}\ul{9}}=-\frac1R.
 \eea
Notice that the components of the spin connections in $N(1, 1)$ only depend on  coordinate $\alpha$.

\end{document}